\documentclass[11pt]{article}
\usepackage{wrapfig}
\usepackage{epsfig}
\usepackage{subfigure}

\begin{document}
\title{Computation of Electrostatic and Gravitational Sag \\
in MultiWire Chambers}

\author{N. Majumdar, S. Mukhopadhyay}
\date{INO Section, Saha Institute of Nuclear Physics\\
1/AF Bidhannagar, Kolkata - 700064\\
\small{nayana.majumdar@saha.ac.in, supratik.mukhopadhyay@saha.ac.in}}
\maketitle

\begin{abstract}
A numerical method of determining the wire sag in a multiwire 
proportional chamber
used in RICH \cite{ALICE} by solving the second order
differential equation which governs the wire stability has been presented.
The three point Finite Difference Method (FDM) has generated a 
tridiagonal matrix equation relating the deflection of wire segments
to the force acting on it. 
The precise estimates of electrostatic force has been obtained from
accurate field computation using a nearly exact Boundary Element 
Method (neBEM) solver \cite{Majumdar06}.
\end{abstract}

\section{Introduction}
The dimension of the multiwire chambers deployed in modern high energy
physics experiments is usually large conforming to the scale of 
experimental setup. The electrostatic instability in such chambers may be 
crucial when the amplitude of the oscillation caused by the action of 
electrostatic
force alone or combined with the gravity becomes comparable to
the electrode spacings.
The study of the wire deflection in such a geometry is usually a complex
affair since an interplay between several physical forces determines
the wire stability. The approximation of constant
or linear dependence of the force on the wire deflection is not adequate 
to solve for the differential equation governing the wire dynamics because
all the wires in the chamber
move in a collective way influencing each other giving rise to a 
nonlinear effect. Since the exact solutions for the differential equation 
involving the nonlinear force are no longer known, it has to be solved
numerically.

Of various methods of estimating the electrostatic sag from the 
differential equation, only the linear and iterative methods have been
attempted in several geometries 
\cite{Majewski96,Vavra97}. In these works, 
the electrostatic force has been estimated
from the 2D field calculation \cite{Garfield} which differs significantly 
from 3D solutions. Owing to the 2D nature of the problem, the sag is
normally overestimated due to the fact that the whole length of the wire 
is considered to be at maximum sag.
In this work, an accurate 3d computation of electrostatic field has been
carried out through
the use of a nearly exact Boundary Element Method (neBEM)  
\cite{Majumdar06} which has yielded precise force estimation.
In order to reduce complexity, only the normal component of the field 
has been considered in the calculation. The deflection
of each segment has been assumed to be very small in comparison to its 
length.

\section{Geometry}
The calculation has been carried out for a geometry similar to that of
RICH detector in ALICE \cite{ALICE}. The anode plane
consists of gold-tungsten wires with $20 \mu $m diameter with 
pitch $4.0$ mm. The upper cathode plane is made of copper-berrylium
wires with diameter $100 \mu$m and pitch $2.0$ mm while the lower one
is a uniform conducting plate.
The separation of upper and lower cathodes from the anode are respectively
$1.68$ mm and $1.83$ mm and length of the detector in Z-direction is
$136.0$cm. The anode plane is supplied with high voltage w.r.t. the cathode
planes.

\section{Numerical Approach}
The second order differential equation in an equilibrium state of the wire
can be written as
\begin{equation}
\label{eqn:DEQ}
T {d^2 y \over{dz^2}} + F_e + F_g = 0
\end{equation}
where $F_e$, $F_g$ are the electrostatic and gravitational forces per unit
length while $T$ the stringing tension of the wire.
Using three point finite difference formula, it can be rewritten as
\begin{equation}
\label{eqn:FDM}
y_{n+1} - 2 y_n + y_{n-1} = - {1 \over T} [F_{en} + F_{gn}].(\delta z)^2
\end{equation}
where $y_n$, $y_{n+1}$ and
$y_{n-1}$ represent the deflections of respective segments. The electrostatic
force on the $n$-th segment has been computed using neBEM solver for the 
given 3D geometry. The required sag due to the action of either of the 
electrostatic and gravitational forces or combined may be obtained from
this equation.
Thus the set of equations for the segments on a wire can be represented as
\begin{equation}
\label{eqn:MatEq}
\textbf{A} \cdot y_n = F_n
\end{equation}
where $\textbf{A}$ is the tridiagonal coefficient matrix whose inverse has been
calculated following standard numerical receipe. In the
present work, five anode wires have been considered with discretization of 
$30$ linear segments while that of the cathode plate has been 
$25 \times 30$. It should be noted that no
plates on the sides of the chamber have been taken into account.

\section{Results and discussions}
\begin{figure}[htb]
\begin{center}
\includegraphics[width=0.45\textwidth]{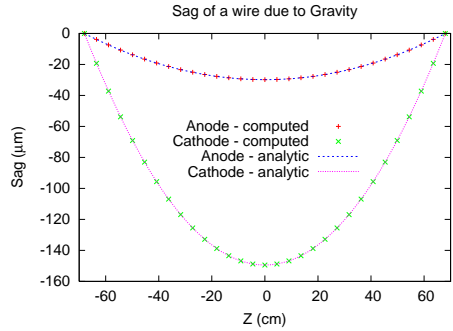}
\caption{\label{fig:GravSagAndCath} Gravitational sag of central anode and
cathode wires}
\end{center}
\end{figure}

The calculation procedure has been validated by calculating wire sag
due to gravitational force and comparing with the analytic solution
for gravitational force only as
\begin{equation}
\label{eqn:Soln}
y(z) = {F_g \over 2T} ({L^2 \over 4} - z^2)
\end{equation}
where $F_g = g\rho\pi r^2$ and $L$, $r$, $\rho$ are the length, radius and 
density of the wire respectively.
The results has been illustrtaed in fig.\ref{fig:GravSagAndCath} which has 
demonstrated the validity of the
method.

\begin{figure}[h]
\centering
\subfigure{\label{fig:EFAnodeWire}\includegraphics[width=0.45\textwidth]
{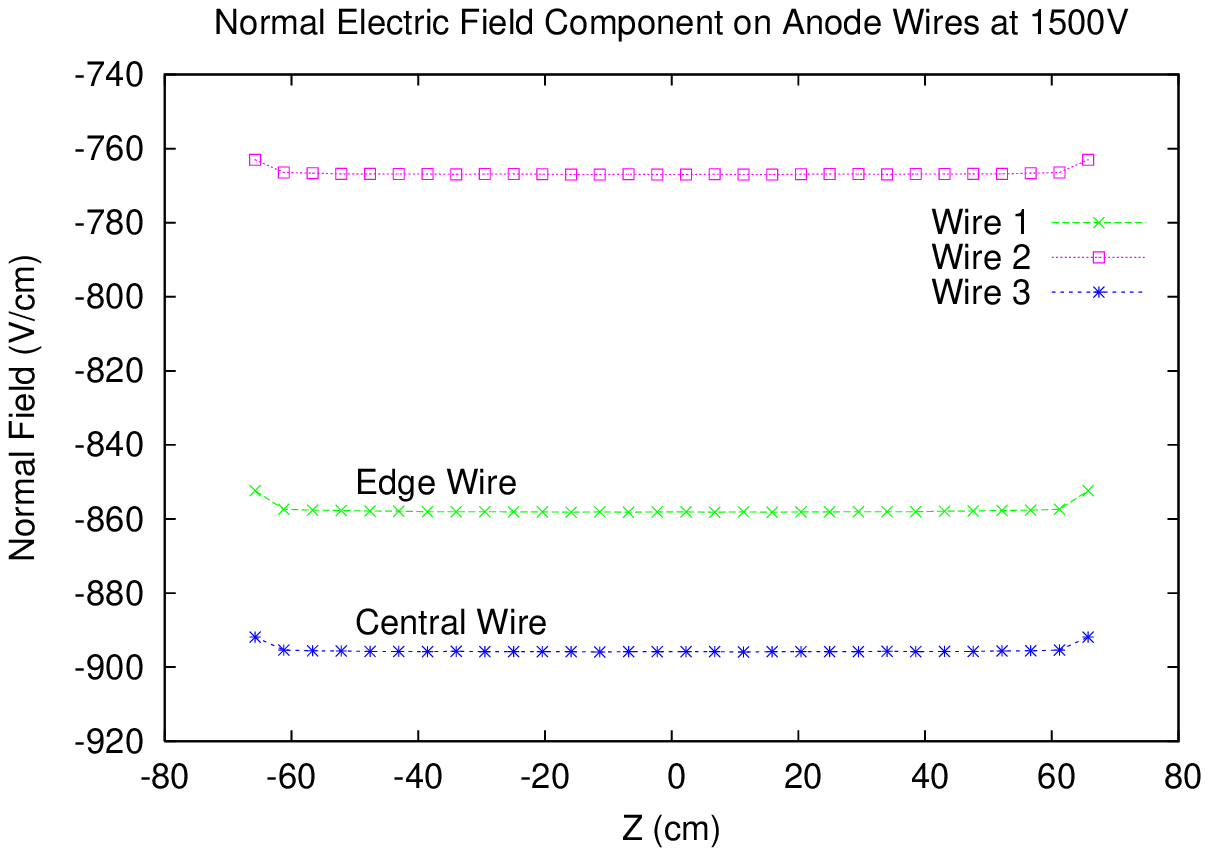}}
\subfigure{\label{fig:EFcathodeWire}\includegraphics[width=0.45\textwidth]
{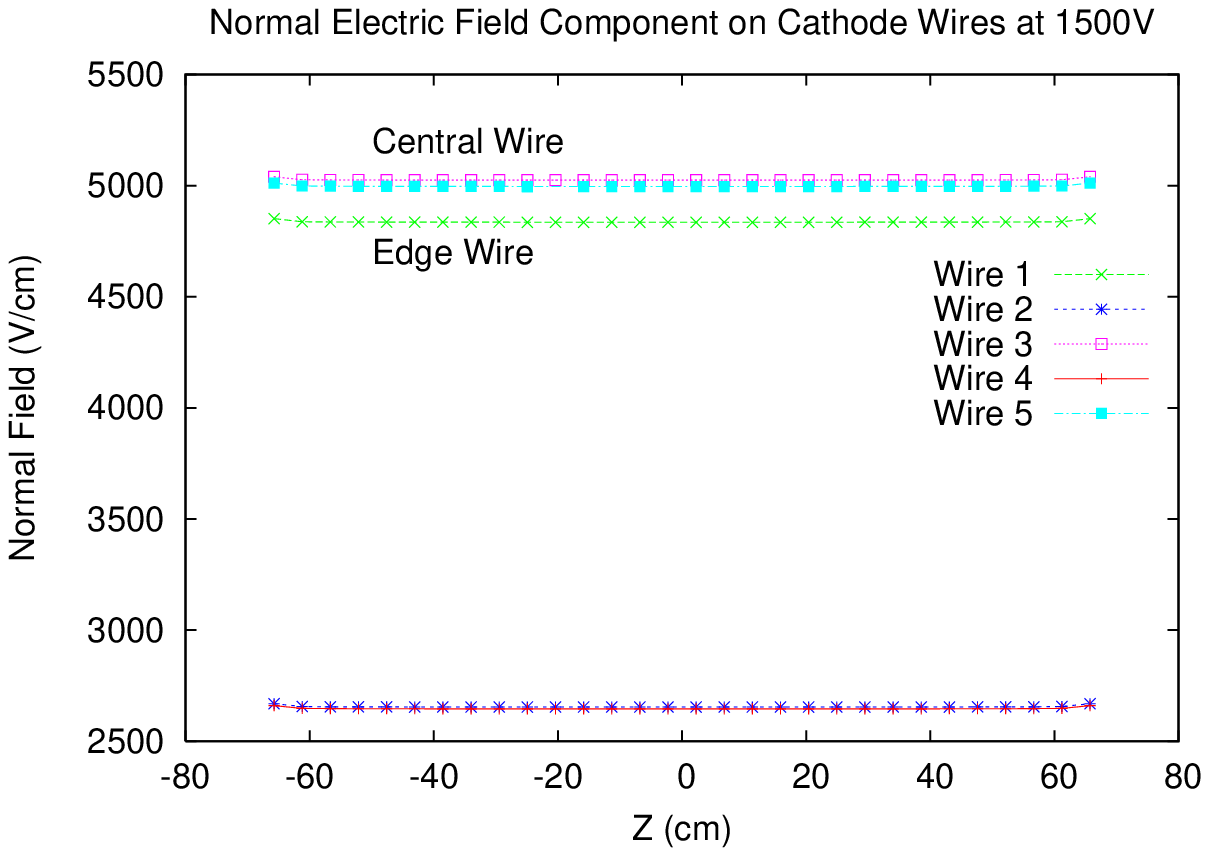}}
\caption{Normal electric field component acting on 
(a) anode wires and (b) cathode wires at $1500$V}
\label{fig:NormalEF}
\end{figure}

The normal electric field components acting on the anode and cathode
wire segments for anode voltage of $1500$V have
been plotted in fig.\ref{fig:NormalEF}. The field component on each 
segment has been calculated from the vectorial addition of field components
at four radial locations on the segment periphery.

\begin{wrapfigure}[15]{r}{0.5\textwidth}
\begin{center}
\includegraphics[width=0.45\textwidth]{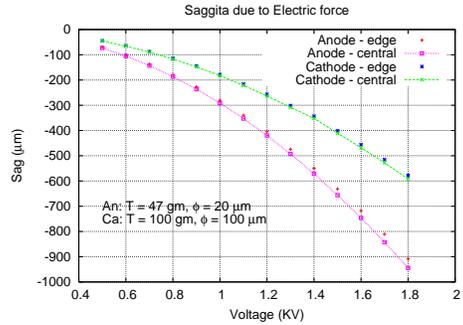}
\end{center}
\caption{\label{fig:WireSag} Wire sags on anode and cathode wires as
a function of anode voltage}
\end{wrapfigure}

The wire sag at the centre due to electrostatic force 
following the solution of tridiagonal matrix equation 
[Eqn.\ref{eqn:MatEq}] has been shown as a function of anode voltage 
in fig.\ref{fig:WireSag} for anode and cathode wires separately.
It is evident from the result that the sag in the anode wire changes
more rapidly than the cathode wires. 
The central wire in the anode plane
has been found to undergo more deflection in comparison to the edge
wires. The calculation of \cite{Majewski96} for wire sags in 
this chamber has reported less deflection in comparison to our result.
In \cite{Majewski96}, an additional restoring electrostatic force has
been considered to be operational when the wire gets deflected which
in turn
has helped to reduce the wire sag. In our calculation, no such
dynamic consideration of the electrostatic force
with the wire deflection has been incorporated. To reproduce the
actual wire sags, an iterative process can be
carried out each time calculating the electrostatic force due to new
position of the deflected wire.

\section{Conclusion}
Using the neBEM solver, the electrostatic field could be accurately calculated
for the three dimensional geometry of multiwire RICH chamber.
An FDM approach to compute the wire sag has been
developed and validated for the case of gravitational sag calculation.
In the
present calculation, no restoring effect of electrostatic force has been 
considered unlike the earlier work which has led to larger sag estimates. 
The restoring force aspect will be implemented in future
by iterative technique to estimate a realistic wire sag in this chamber.


\begin{thebibliography}{10}

\bibitem{ALICE}
ALICE Technical Proposal, CERN-LHCC/95-71, LHCC/P3

\bibitem{Majumdar06}
N.Majumdar, S.Mukhopadhyay,
Nucl. Instr. Meth. Phys. Research, \textbf{A 566}, p.489 (2006)

\bibitem{Majewski96}
P. Majewski,
M.Sc Thesis, Uniwersytet Warszawski, 1996

\bibitem{Vavra97}
J. Va'vra,
SLAC-PUB-7627, August 1997

\bibitem{Garfield}
http://garfield.web.cern.ch/garfield

\end{thebibliography}
\end{document}